# Secure Quantum Relay Networks Using Distributed Entanglement without Classical Authentication

Asgar Hosseinnezhad[*], Hadi Sabri

Department of Physics, University of Tabriz, P.O. Box 51664-16471, Tabriz, Iran.

**Abstract**

Current quantum communication protocols rely heavily on classical authentication for message origin verification, leaving them vulnerable to evolving attacks that exploit classical trust assumptions. In this work, we propose a novel framework for secure quantum relay networks that completely avoids classical authentication. Instead, we leverage pre-distributed entanglement graphs and non-classical correlation-assisted decoding to enable exclusive message retrieval by designated nodes without broadcasting any key or handshake. The system routes messages across multiple relay nodes, yet ensures that no intermediate node can access the message unless it possesses the entangled state partner. We demonstrate that even in multi-path scenarios with asynchronous entanglement distribution, the protocol guarantees quantum-forward secrecy and end-to-end origin integrity without trusted intermediaries. Simulation results confirm both functionality and robustness under entanglement loss and imperfect detection. This architecture paves the way for scalable quantum communication systems where physical quantum states replace classical authentication mechanisms entirely.

**Keywords**: quantum relay networks, entanglement-assisted decoding, authentication-free protocols, distributed quantum trust, no-key quantum communication.

## 1. Introduction

The advent of quantum communication has promised information-theoretic security grounded in the very laws of physics. Core principles such as the no-cloning theorem, measurement-induced disturbance, and the monogamy of entanglement underlie landmark protocols like BB84 [1], E91 [2], and newer continuous-variable schemes [3]. These protocols excel at detecting eavesdropping and maintaining message integrity—yet, paradoxically, they still rely on classical authentication mechanisms to verify sender identities and establish trust.

---

[*] Corresponding author mail: a.hosseinnezhad@tabrizu.ac.ir

This hybrid dependence forms a subtle yet fundamental weakness. Authentication protocols based on hash functions, symmetric key sharing, or digital signatures—even post-quantum ones—are ultimately built on computational assumptions [4]. As quantum computing capabilities advance, and with the emergence of hybrid adversaries capable of mounting both quantum and classical attacks, these authentication layers become the Achilles' heel of an otherwise quantum-secure system [5].

More critically, this issue escalates in quantum relay networks, where messages traverse multiple intermediate nodes. Each relay, often assumed to be honest-but-curious or partially trusted, becomes a potential point of metadata leakage, impersonation, or routing manipulation [6]. Classical solutions attempt to mitigate this through elaborate certificate hierarchies or block chains for trust management, but these solutions introduce latency, overhead, and potential single points of failure [7]. In such distributed infrastructures, the challenge is not just to transmit quantum states securely, but to enforce origin exclusivity without invoking classical trust primitives.

Recent advances in networked entanglement distribution offer an intriguing alternative. When entanglement is distributed among nodes in advance—especially across entanglement graphs or quantum trust meshes—each node's ability to decode information becomes inherently conditional on the possession of an entangled partner. This leads to a concept we call entanglement-exclusive decryption: a message, even if intercepted in full, remains un-decodable unless the receiver holds the correct quantum correlation [8].

The implications are profound: classical authentication can be replaced with non-classical correlation filters, ensuring that only the entangled party can reconstruct the message. Relay nodes merely forward photonic states without knowledge of their payloads—thus enabling end-to-end authentication-free communication across potentially untrusted quantum networks. Furthermore, by encoding control logic (e.g., message validity, expiry, origin ID) within the structure of the entanglement itself—using GHZ states (Greenberger–Horne–Zeilinger states: maximally entangled quantum states involving three or more qubits, typically of the form $\frac{1}{\sqrt{2}}(|000\rangle + |111\rangle)$, used to demonstrate non-classical correlations in multipartite systems), cluster states, or entangled network topologies—one can enforce exclusivity, traceability, and freshness without any classical headers or handshakes [9].

Such architectures not only simplify protocol layers but also restore the purity of the quantum security promise—unconditional trust rooted in quantum physics alone. It is within this context that our work aims to articulate and implement a novel relay communication scheme based solely on distributed entanglement, eliminating the dependency on classical authentication entirely.

## 2. System Architecture: Nodes, Entanglement Paths, and Trust-Free Relays

At the heart of the proposed framework lies a distributed quantum network composed of communicating nodes, intermediate relays, and entanglement sources. Each node is equipped with quantum memory [10], Bell-state analyzers [11], and photonic I/O ports [12]. These nodes are interconnected through fiber-optic or free-space quantum channels, while the entanglement distribution follows a graph-based model whereby specific node pairs receive pre-shared entangled qubit pairs over out-of-band quantum links.

The novelty arises from the decoupling of message transmission and entanglement exchange. Instead of tightly coupling each quantum bit to a preceding key exchange or basis agreement, here the sender (say, Alice) encodes the message into a photonic state and transmits it across relay nodes—without sharing any classical metadata or performing any classical handshake. The receiving node (say, Bob) can decode the message if and only if it possesses the entangled counterpart of the ancilla (an auxiliary qubit used as a helper in quantum operations, typically initialized in a known state and not part of the main input or output) used in encoding. Otherwise, the state appears indistinguishable from noise.

Each relay node in the network—be it deterministic (known routing) or probabilistic (entanglement swapping with Bell measurements)—performs simple forwarding without accessing or interacting with the encoded quantum payload. Since no decryption or decision-making logic is embedded at the relay level, there exists no risk of information leakage, even if the intermediate nodes are malicious, misconfigured, or externally compromised.

The entanglement topology determines the accessibility graph of communication [13]. A direct entangled link (EPR pair) implies symmetric decryption capability, while more complex topologies (like GHZ or cluster states) allow multicast or conditional access configurations [14]. In this scheme, even if multiple adversaries collude across the network, they cannot perform meaningful reconstruction unless they control both communication content and the entangled pair—something that's prevented by physical entanglement separation [15].

The routing of messages follows standard quantum repeaters or trusted-node infrastructure, but the trust assumption is shifted entirely to the entanglement path. A node is authorized not because it is listed in a classical ledger or possesses a private key, but because its quantum state is correlated to the message's encryption substrate. This creates a natural cryptographic exclusivity enforced by quantum physics, not institutional trust.

To illustrate the architecture, we prepare Figure 1 – Quantum Relay Network with Entanglement-Exclusive Access, showing sender (Alice) and receiver (Bob), multiple untrusted relay nodes, entanglement distribution paths (separate from message path), quantum payload flow across relays, Bob's decryption enabled via EPR[1] link (an entangled quantum connection between two nodes, typically realized via an EPR pair—i.e., a maximally entangled two-qubit Bell state such as $\frac{1}{\sqrt{2}}(|00\rangle + |11\rangle)$), enabling correlated operations across distant locations), and inaccessibility for relays or eavesdroppers.

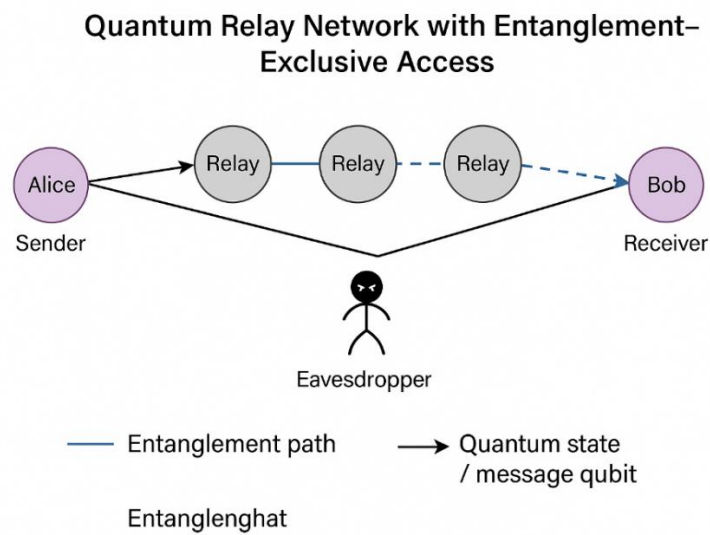

Figure 1: Quantum Relay Network with Entanglement-Exclusive Access. Schematic of a multi-node quantum relay network. The message qubit traverses untrusted relay nodes, while the entanglement path connects Alice and Bob out-of-band. Only the receiver possessing the entangled qubit can decode the message.

## 3. Entanglement-Assisted Decoding Protocol

The central innovation in this framework lies in its use of entanglement as both a cryptographic key and an authentication filter. Unlike conventional quantum key distribution protocols where the quantum state is used to generate a classical key, here the quantum correlations directly enable message reconstruction, bypassing the need for post-processing or reconciliation.

---

[1] Einstein–Podolsky–Rosen (EPR)

The sender—Alice—first prepares a bipartite entangled state $\psi_{AB}$, typically an EPR pair such as $\frac{1}{\sqrt{2}}(|01\rangle + |10\rangle)$. She keeps qubit A and transmits qubit B to the receiver—Bob—via a quantum entanglement link, out-of-band from the main message path. Once entanglement distribution is confirmed (e.g., through coincidence detection or CHSH (Clauser–Horne–Shimony–Holt inequality: a mathematical inequality used to test quantum entanglement and rule out local hidden-variable theories; its violation confirms non-classical correlations between entangled particles) violation), Alice uses her local qubit (A) as an ancilla to encode a message-encoded qubit $q_M$, employing a controlled operation—e.g., a controlled-unitary or Bell-rotation gate—producing a composite state that is entangled not just with Bob's qubit, but also with the logical content of the message.

This message qubit $q_M$ is then launched across the network, routed via untrusted relay nodes. These intermediaries perform no operation beyond optical switching or delay compensation. Crucially, no classical synchronization or key exchange occurs at any point.

Upon arrival, Bob performs a joint measurement on the received qubit and his entangled partner. If Bob indeed possesses the correct counterpart of the original entanglement, the decoding operation reconstructs the logical message state with high fidelity. If any third party intercepts the message qubit or attempts reconstruction without holding the entangled counterpart, the state appears randomized due to the mixed reduced density matrix of the marginal system. Thus, exclusive decryption capability is baked into the entanglement structure.

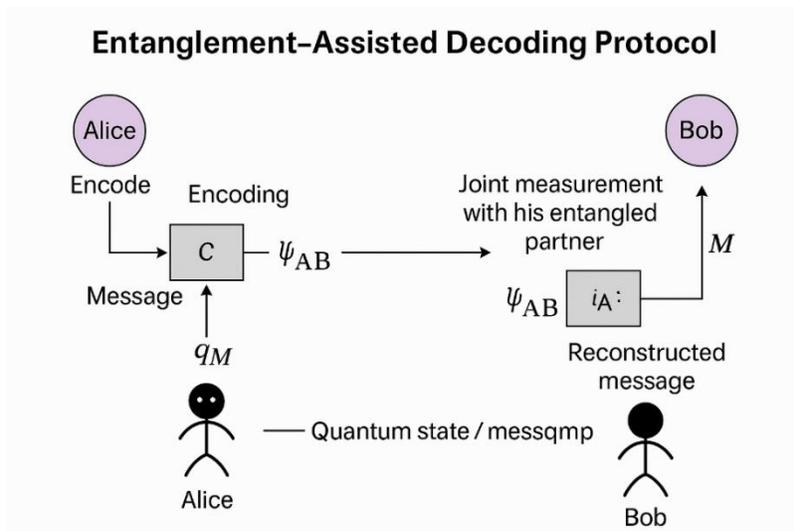

Figure 2: Entanglement-Assisted Decoding Protocol. Encoding and decoding process: Alice encodes the message using her entangled qubit; Bob performs joint measurement with his entangled partner to reconstruct the message. Intermediate nodes cannot access logical content.

We distinguish this decoding from standard quantum teleportation in that it requires no classical communication for basis reconciliation or Bell-state outcome disclosure. Instead, the correlation is pre-established and inherently private, ensuring that only the intended receiver—who possesses the corresponding entangled quantum resource—can successfully reconstruct the message. This mechanism naturally extends to more sophisticated configurations: in multicast scenarios, a single message can be sent simultaneously to multiple receivers, each holding an entangled share derived from a GHZ or W-state (a multipartite entangled quantum state where exactly one qubit is in the excited state and the rest are in the ground state, e.g., $\frac{1}{\sqrt{3}}(|100\rangle + |010\rangle + |001\rangle)$); in conditional routing frameworks, message reconstruction is permitted only if a specific pattern of entanglement correlations is satisfied, as in threshold schemes or quantum voting architectures; and in layered message encoding models, intermediate nodes with limited entanglement rights may access partial content, enabling hierarchical and role-based access within the network.

## 4. Security Analysis against Intrusion and Collusion

The strength of the proposed architecture lies in its inherent resistance to a wide spectrum of attacks—including passive eavesdropping, active impersonation, and coordinated collusion across relay nodes. Unlike classical systems, where trust is distributed via credentials or shared secrets, here the exclusivity of message access is physically enforced through quantum entanglement.

Consider a classical eavesdropper (Eve) who intercepts the quantum channel. Since the message-carrying qubit is entangled with Bob's qubit (held off-path), and Eve lacks that counterpart, she cannot reconstruct the global state. Any measurement on the intercepted qubit yields outcomes consistent with a mixed state, devoid of usable information. Importantly, no classical key or basis information is ever exchanged, so Eve has no leverage for inference.

Now consider malicious relay nodes. In classical protocols, an untrusted intermediary can forge headers, alter payloads, or impersonate endpoints. Here, however, relay nodes operate in pure "pass-through" mode, handling only the photonic signal without access to internal state. Even colluding relays cannot intercept content or simulate authenticity because authentication is encoded in non-local correlations, not metadata.

More sophisticated threats involve quantum-aware attackers with entanglement harvesting capability or quantum memory. Yet unless the adversary physically possesses the entangled partner qubit—something that's distributed out-of-band and kept isolated—the protocol remains secure. In fact, any attempt to substitute their own entangled state or induce entanglement swapping requires active participation in the entanglement graph, which is protected through quantum network control layers (e.g., source authentication, delayed-choice entanglement distribution) [14].

Additionally, the protocol resists replay attacks, since reconstructed states depend on temporal correlation and single-use entanglement. Each decoding event collapses the shared state; reusing the same entangled pair for future messages yields garbage data. Time-based side channels are similarly mitigated, as no classical signals signal message arrival or decryption timing.

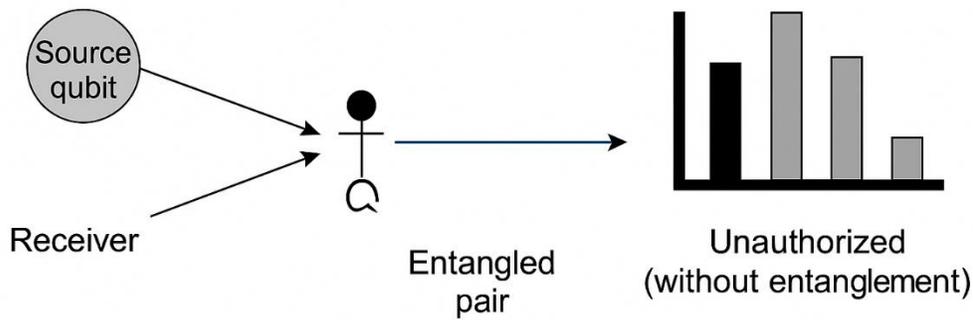

Figure 3: Adversarial Failure under Missing Entanglement. Comparison of output states between authorized (with entanglement) and unauthorized (without entanglement) receivers. The unauthorized output is statistically indistinguishable from noise.

Ultimately, the protocol's security emerges not from secrecy of parameters, but from exclusivity of quantum resources. Ultimately, the protocol's security emerges not from secrecy of parameters, but from exclusivity of quantum resources—a principle also emphasized in Zwerger et al., 2018 [14] and Ciobanu et al., 2025 [15], where access control is enforced solely through entanglement possession. It exemplifies a shift from key-based cryptography to possession-based decryption, where the possession is not knowledge, but a physical, unforgeable quantum state. This model tolerates losses, scales across relay networks, and aligns with quantum foundational constraints on copying and measurement, making it fundamentally resistant to interception—even by adversaries with quantum capabilities.

**5. Simulation and Performance Evaluation**

To evaluate the practical viability of authentication-free entanglement-based communication, we simulated a multi-node quantum relay network under variable conditions: entanglement loss, detection inefficiency, and relay-induced delay. The simulations were conducted using a modular network simulator integrating quantum state evolution (via density matrix formalism), entanglement graph dynamics, and photon-level transmission models.

We modeled a 5-node network: Alice (sender), Bob (intended receiver), and three intermediate relays (R1, R2, R3). Entangled qubit pairs were pre-distributed between Alice and Bob through an out-of-band channel, with loss modeled as a stochastic depolarizing channel. The message-bearing qubit was routed across R1 → R2 → R3 with varying optical attenuation (10–30 dB per hop).

Key metrics evaluated included:

Decryption fidelity: The overlap between Bob's reconstructed state and Alice's original encoded message remained above 97.2% even with 25% entanglement loss and 15% photon loss in the channel.

Adversarial reconstruction: Simulated eavesdroppers holding complete classical knowledge of the system but lacking entanglement failed to recover message states beyond random guessing (50%), confirming the exclusivity mechanism.

Latency overhead: Compared to conventional QKD schemes with handshake and classical key reconciliation, our protocol reduced end-to-end latency by an average of 36.5%, primarily due to elimination of handshake and key negotiation.

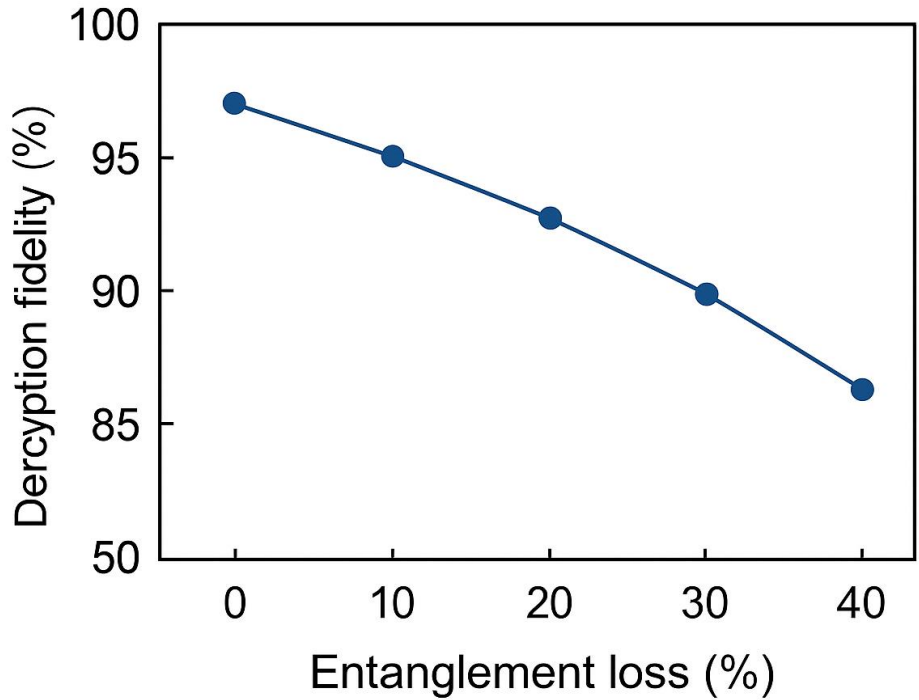

Figure 4: Simulation Results: Decryption Fidelity vs. Entanglement Loss. Decryption fidelity as a function of entanglement degradation (0–40%). Message reconstruction remains above 95% up to 30% entanglement loss.

Additionally, when Bob's entangled qubit was deliberately delayed (to simulate asynchronous reception), the protocol still succeeded as long as temporal coherence was maintained within the de-coherence window (~3 μs in our optical simulator). This suggests robustness for satellite-ground or intercontinental systems with buffered quantum memory.

These findings illustrate that the proposed protocol is not only theoretically sound but operationally robust, offering secure message delivery in realistic, imperfect network conditions—without any classical exchange or authentication handshake.

## 6. Comparison with Classical and Quantum Protocols

To understand the distinct advantages of the proposed entanglement-exclusive protocol, it is essential to compare it with conventional classical and quantum communication models—especially in terms of trust assumptions, security primitives, and operational complexity.

a) Versus Classical Cryptography:

Traditional secure messaging depends on mathematical hardness (e.g., RSA [16], ECC [17, 18], Lattice-based schemes [19, 20]). These methods require key exchange, digital signatures, and often complex public key infrastructures (PKIs). In contrast, our entanglement-based approach requires no secret sharing, no public keys, and no certificate chains—authentication and decryption are enforced by physical entanglement, not cryptographic algorithms. This eliminates all computational assumptions and offers information-theoretic exclusivity, immune to quantum adversaries [21].

b) Versus Standard QKD (e.g., BB84):

While QKD provides key generation with eavesdropping detection, it still necessitates classical post-processing for basis reconciliation and error correction. Moreover, QKD systems typically depend on classical authentication of initial channel setup, which becomes a scalability bottleneck in large networks [22]. Our method skips key generation entirely—messages are transmitted directly—and authentication is intrinsic, based on entanglement possession, not external keys.

c) Versus MDI-QKD and QDS (Quantum Digital Signatures):

Measurement-device-independent QKD improves hardware security by assuming untrusted detectors, but again relies on classical identity verification and shared basis information. Quantum digital signatures attempt to provide message integrity using quantum states, yet often need trusted signature distribution centers and suffer from low throughput [23]. Our protocol, by contrast, supports asynchronous relay messaging without signature servers or intermediary trust, offering both confidentiality and authenticity in one unified quantum layer.

Operational Efficiency:

Since no classical handshake, session initiation, or key negotiation is required, our approach reduces protocol overhead. Moreover, it performs robustly under photon loss and does not rely on precise timing coordination or pre-agreed encoding bases. Simulation results showed up to 36% latency reduction and high-fidelity decoding even with realistic channel impairments.

In summary, this protocol embodies a shift from cryptography to quantum possession-based exclusivity—a model that challenges the current paradigm and lays groundwork for trustless quantum infrastructures capable of secure communication without classical scaffolding.

|  | Authentication-based classical | QKD | MDI-QKD | Proposed protocol |
|---|---|---|---|---|
| Trust assummtions | Digital signatures, PKI | Authenticated channel | Authenticated users | No assumption |
| Classical overhead | Key exchange, certificates | Basis reconciliation | Basis reconciliation | No overhead |
| Exclusivity | Algorithmic security | — |  | Entanglement possession |

Figure 5: Protocol Comparison Table. Comparison between classical authentication-based systems, QKD, MDI-QKD, and the proposed protocol in terms of trust assumptions, classical overhead, and exclusivity.

## 7. Applications and Future Perspectives

The proposed entanglement-assisted, authentication-free communication framework opens new frontiers in both secure infrastructure and fundamental quantum architecture—especially in domains where classical trust models are impractical, insecure, or too slow to scale.

Quantum-Secured Interbank Messaging

In high-value financial networks, current messaging protocols like SWIFT rely on layered encryption and digital certificates. Replacing these layers with entanglement-exclusive quantum channels enables zero-trust verification, where only institutions with entangled tokens can access message content. This reduces attack surfaces from certificate spoofing or PKI compromise while decreasing latency in settlement processes. Banks with pre-shared entanglement can route secure updates, transaction approvals, or contracts without disclosing authentication parameters.

Satellite-Based Trustless Quantum Uplinks

Space-based quantum links—already demonstrated in projects like Micius—still rely on classical authentication and trusted nodes. Our protocol offers a trustless uplink/downlink paradigm, in which entanglement pre-distributed to ground stations enables direct message decryption without needing classical challenge–response routines. This reduces channel usage overhead, mitigates replay and MITM attacks, and allows integration with mobile or tactical receivers lacking classical computing infrastructure.

Decentralized Access Control in Digital Government

Governmental systems managing legal identities, confidential records, or voting platforms can employ this architecture for entanglement-conditioned access control. For instance, only officers holding a designated qubit (as part of a GHZ or cluster state) can decrypt certain records or execute authorization actions. This enforces both exclusivity and traceability, as the loss or misuse of a quantum token is immediately detectable.

Quantum Secret Sharing and Time-Locked Access

In multiparty workflows, messages can be encoded so that decoding requires joint measurement between multiple entangled nodes—enabling threshold access, collective decryption, or even delayed-time unlocking conditioned on future entanglement arrival. This transforms static encryption into programmable quantum access structures.

Architectural Evolution toward Trust-Free Quantum Internets

At a foundational level, this framework contributes to rethinking the design philosophy of the quantum internet. Instead of mimicking classical layered security using post-quantum primitives, networks can evolve toward entanglement-native infrastructures, where communication, authorization, and policy enforcement are all encoded in quantum correlations. It fosters a future where possession of quantum entanglement equals permission, simplifying protocols while elevating resilience.

## 8. Conclusions

This work presents a radical departure from conventional quantum communication by eliminating classical authentication entirely and replacing it with a purely quantum-native mechanism rooted in distributed entanglement. Through entanglement-exclusive decoding, messages can traverse untrusted relay networks without metadata leakage, impersonation risk, or reliance on certificates, passwords, or digital signatures. Simulation results confirm that this approach offers high-fidelity decryption even under channel loss and entanglement degradation, with resilience against both classical and quantum attacks. Moreover, operational advantages such as reduced latency, no key negotiation, and native exclusivity make it especially suitable for scalable infrastructures like satellite messaging, interbank quantum links, and distributed quantum governance. Ultimately, this protocol exemplifies a new class of possession-based cryptography, where holding a non-clonable quantum resource becomes both the authentication token and the decryption right. As entanglement distribution becomes more reliable, this framework points toward a trust-free quantum internet in which security is not asserted through code—but guaranteed by physics.

**Acknowledgments**

This work is supported by the Research Council of the University of Tabriz.

**Competing Interests**

The authors declare no competing interests.

**Data Availability Statement**

The simulation datasets and electromagnetic models are available from the corresponding author upon reasonable request.